\documentclass[conference]{IEEEtran}
\pdfoutput=1
\IEEEoverridecommandlockouts

\usepackage{cite}
\usepackage{amsmath,amssymb,amsfonts}
\usepackage{algorithmic}
\usepackage{graphicx}
\usepackage{textcomp}
\usepackage{xcolor}
\def\BibTeX{{\rm B\kern-.05em{\sc i\kern-.025em b}\kern-.08em
    T\kern-.1667em\lower.7ex\hbox{E}\kern-.125emX}}

\usepackage{graphicx}
\usepackage{subfigure}
\graphicspath{{figures/}}
\usepackage{amsmath}
\usepackage{amssymb}
\usepackage{xcolor}
\usepackage{siunitx}
\usepackage{amsthm}
\usepackage{multicol}
\usepackage{lipsum}
\usepackage{tikz}

\newcommand{\hBS}{h_{\text{BS}}}
\newcommand{\hB}{h_{\text{B}}}
\newcommand{\hRIS}{h_{\text{RIS}}}
\newcommand{\xRIS}{x_{\text{RIS}}}
\newcommand{\yRIS}{y_{\text{RIS}}}
\newcommand{\xB}{x_{\text{B}}}
\newcommand{\yB}{y_{\text{B}}}
\newcommand{\lambdaB}{\lambda_{\text{B}}}
\newcommand{\Gt}{G_{\text{t}}}
\newcommand{\Gr}{G_{\text{r}}}
\newcommand{\dFA}{d_{\text{FA}}}
\newcommand{\ARIS}{A_{\text{RIS}}}
\newcommand{\ABS}{A_{\text{BS}}}
\newcommand{\Ano}{A_{\text{no}}}
\newcommand{\Aboth}{A_{\text{both}}}
\newcommand{\Rboth}{R_{\text{both}}}
\newcommand{\Rno}{R_{\text{no}}}
\newcommand{\gammath}{\gamma_{\text{th}}}
\newcommand{\rhocover}{\rho_{\text{cover}}}
\newcommand{\Ugrid}{U_{\text{grid}}}
\newcommand{\PLRIS}{\mathsf{PL}_{\text{RIS}}}
\newcommand{\PLBS}{\mathsf{PL}_{\text{BS}}}

\begin{document}

\title{Optimizing the Deployment of Reconfigurable Intelligent Surfaces in MmWave Vehicular Systems\\
}

\author{\IEEEauthorblockN{Xiaowen Tian$^{\dag}$, Nuria Gonzalez-Prelcic$^{\dag}$, Robert W. Heath Jr.$^{\dag}$}
\IEEEauthorblockA{$^{\dag}$\textit{Electrical and Computer Engineering Department} \\
\textit{North Carolina State University}\\
Raleigh, NC, 27606, USA \\
E-mail: \texttt{\{xtian8,  ngprelcic, rwheathjr\}@ncsu.edu}}
}

\maketitle

\begin{abstract}
Millimeter wave (MmWave) systems are vulnerable to blockages, which cause signal drop and link outage. One solution is to deploy reconfigurable intelligent surfaces (RISs) to add a strong non-line-of-sight path from the transmitter to receiver. To achieve the best performance, the location of the deployed RIS should be optimized for a given site, considering the distribution of potential users and possible blockers. In this paper, we find the optimal location, height and downtilt of RIS working in a realistic vehicular scenario. Because of the proximity between the RIS and the vehicles, and the large electrical size of the RIS, we consider a 3D geometry including the elevation angle and near-field beamforming. We provide results on RIS configuration in terms of both coverage ratio and area-averaged rate. We find that the optimized RIS improves the average averaged rate fifty percent over the case without a RIS, as well as further improvements in the coverage ratio.
\end{abstract}

\begin{IEEEkeywords}
Reconfigurable intelligent surfaces, millimeter wave, blockages, deployment, vehicular systems
\end{IEEEkeywords}

%
\IEEEpeerreviewmaketitle

\begin{figure*}[t]
  \centering
  \subfigure[{System model.}]{
  \label{fig:systemmodel}
  \includegraphics[width= 4.4 in]{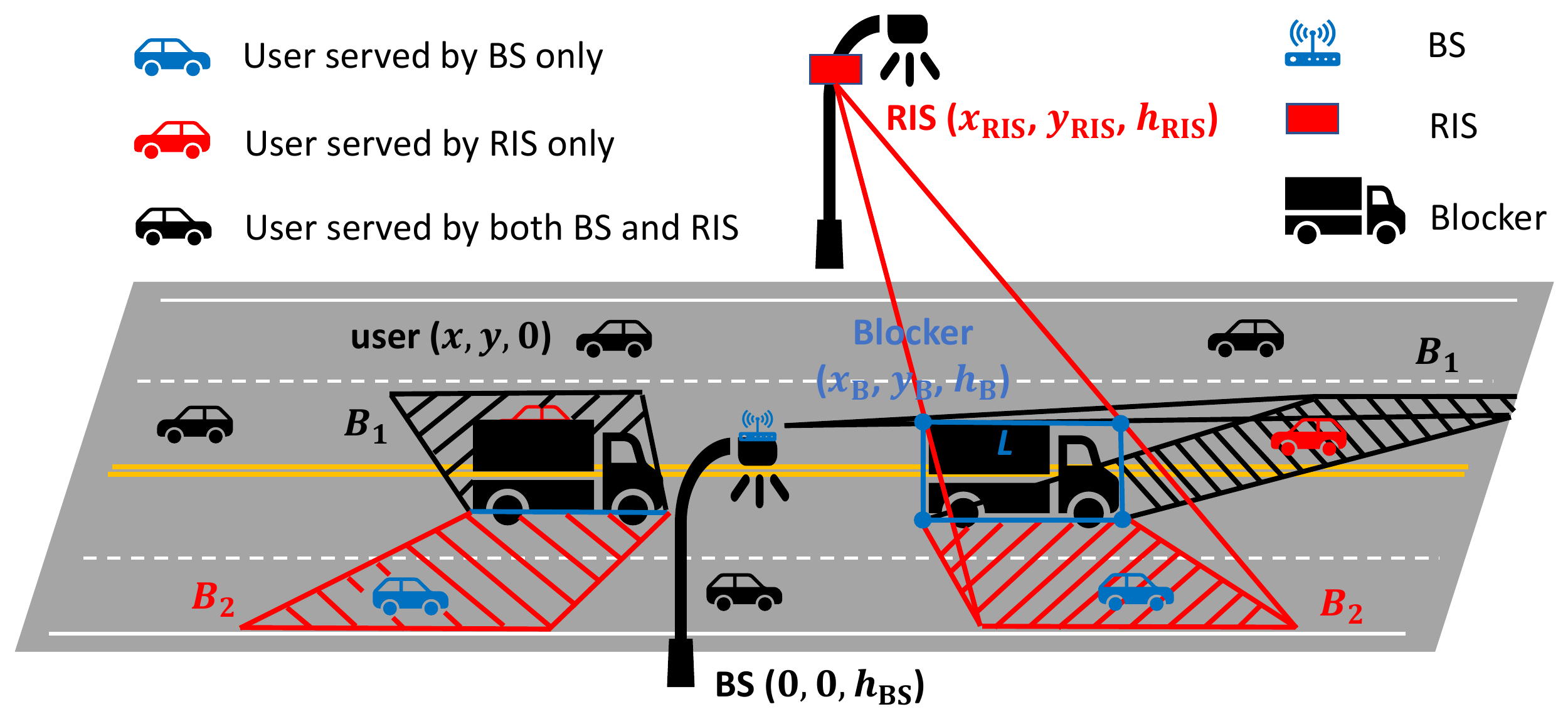}}
  \hspace{0em}
  \subfigure[RIS model.]{
  \label{fig:RIS model}
  \includegraphics[width= 2.2 in]{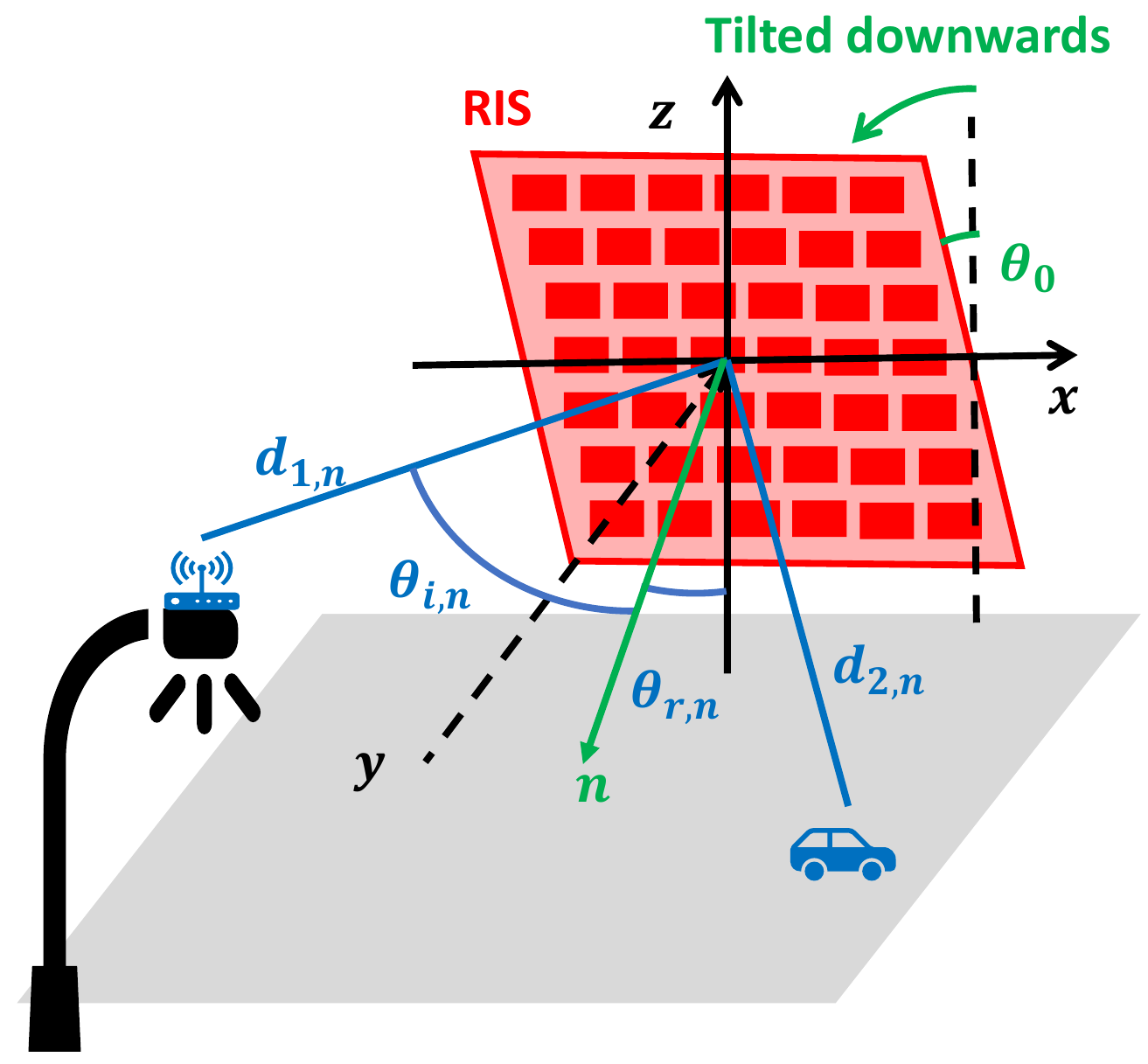}}
  \vspace{-0.1cm}
  \caption{ (a) A RIS-assisted mmWave vehicular system model. The red users are only served by RIS and the blue users are only served by BS because of blockages. The black users are served by both BS and RIS. The black shaded region $B_1$ can only be served by RIS and the red shaded region $B_2$ can only be served by BS.(b) Parameters illustration for a RIS working in the near-field beamforming model. It is tilted downwards to better serve the users who are all located on the ground. }
  \label{fig:system}
  \vspace{-0.2cm}
\end{figure*}

\section{Introduction}

Reconfigurable intelligent surfaces (RIS) provide an alternative solution to relays to reduce the impact of blockage in millimeter wave networks \cite{RIS}.
Extensive recent work has focused on designing strategies to configure a RIS under different metrics \cite{WuZhangIntelligentReflectingSurfaceEnhanced2019}, or to estimate the corresponding composite channel \cite{HuEtAlTwoTimescaleChannelEstimationReconfigurable2021}.
Although the location of the RIS heavily impacts its ability to overcome blockage, limited work has been devoted to analyzing and optimizing RIS deployment.

The optimization of the RIS placement was first analyzed in \cite{ZengEtAlReconfigurableIntelligentSurfaceRIS2020}, where a RIS is used to extend the coverage in a cellular setting with one base station (BS) and one user. For this particular configuration, the optimal RIS placement was close to the cell edge. The optimal location of a large RIS deployed on the facade of a building was investigated in \cite{NtontinEtAlReconfigurableIntelligentSurfaceOptimal2021}, with the RIS partially illuminated due to its large aperture. The optimal BS-RIS horizontal location was obtained to maximize the end-to-end signal-to-noise ratio (SNR).
Instead of placing RIS vertically on the building facade, the work in
 \cite{LuEtAlAerialIntelligentReflectingSurface2021} considers a RIS mounted on aerial platforms, with the surface parallel to the ground.  In this case,  the optimal location and height were obtained to maximize the worst-case SNR. The stochastic behavior of the blockage and users was exploited in \cite{GhatakPlacementIntelligentSurfacesRSSIBased2021,GhatakEtAlWhereDeployReconfigurableIntelligent} to optimize the  RIS location.
In \cite{GhatakPlacementIntelligentSurfacesRSSIBased2021}, the blockage probability was derived under the assumption that both, the BS-RIS-user and BS-user links, can be blocked. The coverage probability for this same setting was computed  in \cite{GhatakEtAlWhereDeployReconfigurableIntelligent}. Using those results, the optimal BS-RIS horizontal distance and height were obtained to minimize the joint blockage probability and maximize the downlink coverage probability.

The previous studies on RIS placement make certain simplify assumptions about the geometry. First, some work like \cite{RIS}-\cite{ZengEtAlReconfigurableIntelligentSurfaceRIS2020},\cite{LuEtAlAerialIntelligentReflectingSurface2021} neglects the elevation angle of the incident and reflected signals at the RIS. This results in neglecting the orientation dependent on the gain, and users might get lower rates than expected \cite{TangEtAlWirelessCommunicationsReconfigurableIntelligent2021}.  Second, other work like \cite{NtontinEtAlReconfigurableIntelligentSurfaceOptimal2021}, \cite{GhatakPlacementIntelligentSurfacesRSSIBased2021} and \cite{GhatakEtAlWhereDeployReconfigurableIntelligent} neglects near-field effects and assume that the RIS is located in the far-field. This results in a miss-alignment loss due to the curvature of the spherical wave \cite{BjornsonEtAlPrimerNearFieldBeamformingArrays2021}.
Vehicular mmWave systems are a promising application of RIS to overcome blockages due to other vehicles, with RIS located on  buildings or infrastructure nearby the roadway. In this case, a 3D model is needed where the gains of RIS include the impact of both the elevation angle and the near-field beamforming effects.

In this paper, we devise the optimal RIS location for a mmWave vehicular communication system with a proximate RIS operating in the near-field beamforming mode. We consider a roadway where a BS is located on one side of the street and a RIS is located on the other side of the street and tilted downwards. Cars may be located anywhere along the roadway with their link to the BS, RIS, or both possibly blocked by randomly located trucks. In this configuration, we derive the optimal RIS configuration in terms of the location along the roadway, height, and downtilt angle to compensate for the near-field effect. Based on the optimal RIS configuration, we derive the coverage ratio and the area-averaged rate as the performance metrics used to optimize the RIS deployment. These metrics are computed by dividing the considered region according to the serving status (served by RIS, by BS, by RIS and BS or unserved) caused by the randomly distributed blockers, and  assuming that both links can be blocked. The optimal horizontal RIS location is obtained as the nearest point to the BS, which is verified by numerical studies, while the optimal RIS height and tilt angle can be obtained through a simple search. To the best of our knowledge, this the first work that optimizes the deployment of RISs considering the near-field effect and randomly distributed blockers under a 3D system model.

\section{System model}

We consider a mmWave vehicular scenario where one BS is placed on one side of the road and one RIS is placed on the opposite side, as illustrated in Fig.~\ref{fig:system}. The RIS is tilted downwards and facing the street.
The vehicles acting as blockers are moving in one lane, and the users are uniformly distributed across the different lanes.
The users can be served by the BS-RIS-user link, the BS-user link, or both.
A target user is located at $(x,y,0)$ and the RIS operates in the near field mode.
The center of the BS array is located at $(0,0,\hBS)$, while the center of the RIS is located at $(\xRIS, \yRIS, \hRIS)$, where $\hBS$ and $\hRIS$ are the heights for the center of BS and RIS.
The RIS is tilted with an angle $\theta_0$ with respect to the vertical.
The vertical region containing the blocker is defined by a rectangle of length $L$ and height $\hB$, with its bottom left corner at $(\xB, \yB, 0)$, as illustrated in Fig. \ref{fig:systemmodel}.
According to the blockage model B recommended in the 3GPP specifications for outdoor vehicles \cite{3GPPblockage}, $L=4.8$ m.
The number of blockers on one lane $M$ is modeled with a Poisson distribution with parameter $\lambdaB$, with each blocker independently, identically  and uniformly  distributed on the lane, i.e.,
\begin{equation}
    P\{M=p+1\} = \frac{\lambdaB^p}{p!} e^{-\lambdaB}, ~p=0, 1, 2,...
    \label{eq:Poisson}
        \vspace{-0.0cm}
\end{equation}
We assume that there is at least one blocker in the system.
Thus, the expectation of $M$ is $\mathbb{E}\{M\} = \lambdaB + 1$.
For simplicity, we also assume that the RIS and the BS are always higher than the blocker, so that the BS-RIS link is never blocked. 

\addtolength{\topmargin}{0.09in}

\subsection{Path loss model and rate}

Let us denote $\Gt$ and $\Gr$ as the antenna gains at the transmitter and receiver.
$a$ and $b$ as the length and width of each RIS element, $\lambda$ as the wavelength, and $N$ as the number of RIS elements.
The geometric parameters are illustrated in Fig. \ref{fig:RIS model}.
Explicitly, $\theta_0$ is the RIS tilt angle, $\theta_{i,n}$ ($\theta_{r,n}$) is the elevation angle of the impinging (reflected) signal at the $n$-th element of the RIS, $d_{1,n}$ and $d_{2,n}$ are the distances from the BS and the user to the $n$-th RIS element, and $F(\theta_n, \phi_n)$ is the radiation function of the $n$-th element of the RIS, which decides the corresponding gain of the $n$-th element of the RIS $G$.
The path loss corresponding to the BS-RIS-user link is then calculated by using the near-field beamforming model as
\begin{small}
\begin{equation}
    \PLRIS = \frac{1}{\Gt \Gr G} \frac{64 \pi^3} {ab \lambda^2} \frac{1}{ \left|\sum_{n=1}^N \frac{\sqrt{F(\theta_{i,n}, \phi_{i,n}) F(\theta_{r,n}, \phi_{r,n})} }{d_{1,n} d_{2,n}} \right|^2}.
    \label{eq:PL_RIS}
\end{equation}
\end{small}
\hspace{-0.3 cm} Note that the RIS tilt angle $\theta_0$ will impact not only the two elevation angles, but also the two distances since the locations of the RIS elements also change with tilting.

To achieve maximum received signal power, the phase shift of the $n$-th RIS element is optimally configured as
\begin{equation}
    \psi_n = \mathrm{mod}\left(\frac{2 \pi (d_{1,n} + d_{2,n})}{\lambda}, 2 \pi \right).
\end{equation}

The near-field and far-field boundary for the RIS are defined from the Fraunhofer array distance \cite{BjornsonEtAlPrimerNearFieldBeamformingArrays2021} as
\begin{equation}
    \dFA = \frac{2 N \sqrt{a^2 + b^2}}{\lambda}.
    \label{eq:dFA}
\end{equation}
Assuming operation at mmWave bands and common dimensions for a vehicular setting as chosen in Sec. \ref{sec:sim}, the $\dFA$ is much longer than both $d_1$ and $d_2$, so the near-field beamforming path loss model in (\ref{eq:PL_RIS}) is realistic.

The path loss for the BS-user link is calculated as
\begin{equation}
   \PLBS = \frac{1}{\Gt \Gr} \left(\frac{4\pi}{\lambda} \right)^2 d^2, \label{eq:PL_BSgeneral}
\end{equation}
where $d$ is the distance of the BS-user link.

Let us denote the regions where users can only be served by the BS-RIS-user link as $\ARIS$, and those served only by the BS as $\ABS$.
Given the corresponding path loss models, the rate at serving regions $\ARIS$ and $\ABS$ can be calculated as
\begin{equation}
    R_{m} = \log_2 \left(1+ \frac{\rho}{\mathsf{PL}_{m}} \right) ,
    \label{eq:rateRISm}
\end{equation}
where $m \in \{\text{RIS}, \text{BS}\}$ and $\rho$ is the SNR at the receiver side.

When both links are available, we assume that the BS knows the perfect channel state information (CSI).
With BS carrying multiple antennas and able to beamforming towards two directions, it splits the transmit power between the BS-RIS-user link and the BS-user link with optimal power allocation ratio $\beta$.
Let us denote the region where users can be served by both BS and RIS as $\Aboth$.
Its corresponding rate is calculated as
    \vspace{-0.0cm}
\begin{equation}
    \Rboth =  \log_2 \left(1+  \rho  \frac{ \beta}{\PLRIS } \right) + \log_2 \left( 1 + \rho \frac{ 1-\beta}{\PLBS} \right).
    \label{eq:rateboth}
\end{equation}
    \vspace{-0.0cm}

Finally, let us denote the region where users can not be served as $\Ano$.
To better separate the serving status at mmWave bands, the path loss in this region is assumed to be $\mathsf{PL}_\text{no} = \infty$, and the corresponding rate is $\Rno = 0$.

\begin{figure}[t]
    \centering
    \vspace{0.0 in}
    \includegraphics[width= 3.1 in]{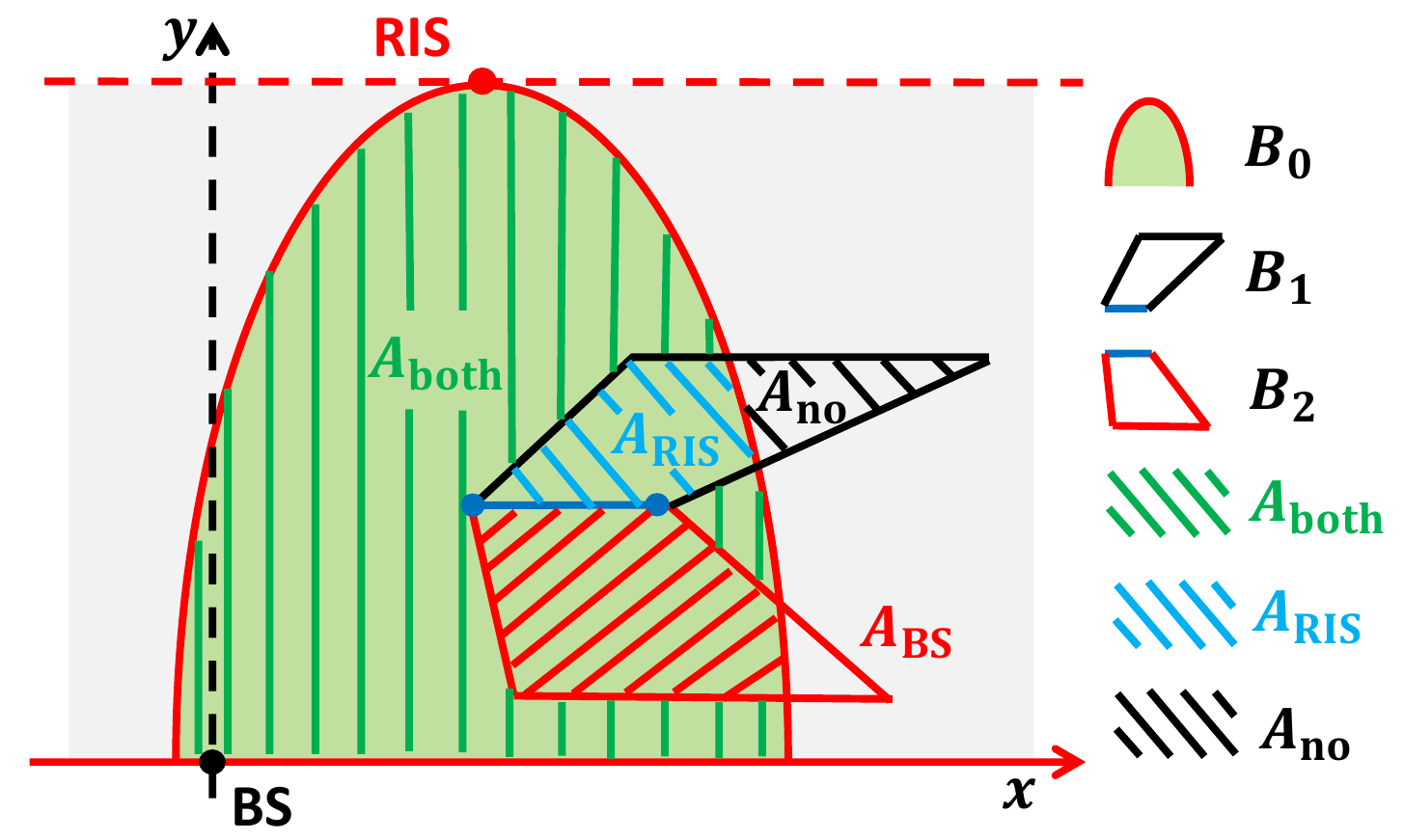}
    \caption{Illustration of different regions. The light green region $B_0$ is the RIS coverage range, $B_1$ and $B_2$ are the shadow regions by the blocker. The green shaded region $\Aboth$ is served by both BS and RIS. The blue shaded region is only served by RIS. The black shaded region $\Ano$ is not covered because it's inside the blocker's shadow and it's beyond the RIS serving range. The region excluding $\Aboth$, $\ARIS$ and $\Ano$ is served only by BS denoted as $\ABS$ and is not shown in the figure. The red shaded region is $B_2 \bigcap B_0$.  }
    \label{fig:fourregions}
    \vspace{-0.3 cm}
\end{figure}

\enlargethispage{-2\baselineskip}

\subsection{Serving regions}

Denote the considered region as $C$.
To account for the differences in serving range of both links, a threshold $\gammath$ is applied to indicate the maximum path loss for supporting the communication.
BS is set to cover the whole target user locations, i.e., $\PLBS \leq \gammath, \forall (x,y) \in C$.
The path loss of the BS-RIS-user link (\ref{eq:PL_RIS}), however, increases more quickly because of the multiplication of distances, resulting in the fact that the RIS may not be able to cover the whole region $C$.
Therefore, there are regions that can not be supported by the BS-RIS-user link.
We denote the region where users can be supported by RIS as $B_0$, with path loss $\PLRIS \leq \gammath, \forall (x,y) \in B_0$.

For now, let us ignore the RIS serving range issue. Due to the existence of the blocker, two shadow areas are generated where the users can only be served by either the BS or the RIS.
Let us define the shadow region that can only be reached by RIS as $B_1$, illustrated as a black shaded area in Fig. \ref{fig:systemmodel}.
Similarly, the region that can only be served by BS is defined as $B_2$, illustrated as a red shaded area in Fig. \ref{fig:systemmodel}.

Now let us consider the RIS serving range together with the shadow regions created by the blockers.
The whole considered region $C$ can be divided into four regions according to the serving status, as illustrated in Fig. \ref{fig:fourregions}:
\begin{itemize}
    \item The region supported by RIS only $\ARIS$.
This region is the intersection of region $B_1$ and $B_0$, i.e., $\ARIS = B_1 \bigcap B_0$.
    \item The unserved region $\Ano$.
This region is within the blocked region $B_1$, and it is due to the fact that the RIS cannot reach that far, i.e., $\Ano = B_1 - B_1 \bigcap B_0$.
    \item The region served by both BS and RIS $\Aboth$.
This region is within RIS coverage $B_0$, and it is outside of the shaded regions $B_1$ and $B_2$, i.e., $\Aboth = B_0 - B_1 \bigcap B_0 - B_2 \bigcap B_0$.
    \item The region served only by the BS $\ABS$.
    This region is the whole considered region minus all three other regions, i.e., $\ABS = C - \Ano - \ARIS - \Aboth = C - B_0 - B_1 + B_1 \bigcap B_0 + B_2 \bigcap B_0$.
\end{itemize}
\vspace{-0.2cm}

\addtolength{\topmargin}{-0.09in}

\addtolength{\topmargin}{0.32in}
\enlargethispage{-2\baselineskip}

\subsection{Area-averaged rate}
\label{sec:rate}

To evaluate the performance in a given region, we define the regional area-averaged rate $\Bar{\rho}_{m}$ as
\vspace{-0.1cm}
\begin{equation}
\vspace{-0.1cm}
    \Bar{\rho}_{m} = \frac{1}{S_{m}} \iint_{A_{m}} R_{m}(x,y) \,dx \,dy,
\end{equation}
where $S_m$ is the area of $A_m$,  $m \in \{\text{RIS}, \text{BS}, \text{both}, \text{no}, C\}$, and $R_m(x,y)$ is the rate at location $(x,y)$.
The area-averaged rate for the whole considered region is
\vspace{-0.2cm}
\begin{equation}
\vspace{-0.1cm}
    \Bar{R} = \frac{1}{S_C}\sum S_m \Bar{\rho}_{m}.
    \label{eq:averagepl}
\end{equation}
It is calculated given the coordinates of the BS, RIS, users and blockers, along with the RIS tilt angle.
\vspace{-0.1cm}

\section{Optimal RIS placement}
\label{sec:placement}

In this section, we will investigate the RIS deployment problem under two metrics, namely, coverage ratio and area-averaged rate.

\vspace{-0.2cm}
\subsection{Definition of Performance Metrics}

\subsubsection{Coverage ratio}

The coverage ratio is defined as the ratio of unserved area with respect to the whole considered area,
\vspace{-0.1cm}
\begin{equation}
    \rhocover = 1- \frac{S_\text{{no}}}{S_C}.
    \label{eq:coverageratio}
\end{equation}

\vspace{-0.1cm}
In our paper, we intend to find the optimal RIS placement $\xRIS^*$, $\hRIS^*$ and $\theta_0^*$ such that the average coverage ratio $\mathbb{E}\{\rhocover\}$ is maximized, where the expectation comes from the fact that the number of blockers $M$ and the locations $\xB$ are random variables.

\subsubsection{Area-averaged rate}
Since maximizing the coverage ratio is equivalent to maximizing the number of users with $\mathsf{PL}(x,y) \leq \gammath$, only the lower bound of the rate performance is guaranteed. 
Therefore, we also consider the optimal RIS deployment that maximizes the area-averaged rate, i.e., $\mathbb{E}\{\Bar{R}\}$.

\subsection{Optimizing the deployment}
Since the serving regions can not be explicitly expressed and it has not been possible to find the closed form solution to the integrals in (\ref{eq:averagepl}), we define a grid to express users' locations within the considered region and numerically compute the performance metrics.
Denote $U_\text{grid}$ as the grid of possible user locations.
Thus, the coverage ratio can be written as
\begin{equation}
\vspace{-0.0cm}
    \rhocover = 1 - \frac{N_\text{no}}{N_{C}},
    \label{eq:rho_N}
    \vspace{-0.0cm}
\end{equation}
where $N_\text{no}$ and $N_{C}$ are the number of grid users in region $\Ano$ and $C$.
The area-averaged rate becomes the average rate, computed as
\begin{equation}
    \Bar{ R} = \frac{1}{N_C} \sum_{(x,y) \in \Ugrid} R_u,
    \label{eq:R_N}
     \vspace{-0.05cm}
\end{equation}
with $R_u$ representing the actual rate for user $u$.

\subsection{Optimal RIS location}
\label{sec:opt_x_RIS}
Note that the elevation angles $\theta_{i,n}$, $\theta_{i,n}$ depend on the distances $d_{1,n}$ and $d_{2,n}$ as $\cos\theta_{i,n} \propto \frac{1}{d_{1,n}}$ and $\cos\theta_{r,n} \propto \frac{1}{d_{2,n}}$.
With this approximation, the path loss in \eqref{eq:PL_RIS} can be rewritten as
\begin{equation}
    \PLRIS \propto (d_{1,n} d_{2,n})^\alpha, \alpha > 1,
\end{equation}
with distances $d_{1,n} \approx d_1$ and $d_{2,n} \approx d_2$. By measuring from the center of RIS $d_1^2 = (\hBS - \hRIS)^2 + \yRIS^2 + \xRIS^2$, and $d_2^2 = \hRIS^2 + (\yRIS-y)^2 + (\xRIS-x)^2$.
Therefore, to have minimum $d_{1,n} d_{2,n}$ and thus maximize the RIS coverage, $\xRIS^* = 0$ is the optimal horizontal location for the RIS.

With the deployment strategy of placing the BS and RIS on opposite sides of the street, blockage will only affect one of the serving links, meaning the outage happens only when the blocked region can not be reached by RIS. Therefore, the coverage ratio should be very close to 1.

\subsection{Optimal RIS height and tilt angle}
\label{sec:opt_h_RIS_theta0}

Intuitively, having a higher RIS will lead to smaller blocked areas, but the path loss will also increase.
In addition, having a tilted RIS is beneficial because it provides a smaller elevation angle for the RIS-user link, but it might be at the cost of creating a larger elevation angle for the BS-RIS link.
Furthermore, having a lower RIS will prohibit a large tilted angle, because the RIS might not be able to ``see" the BS.
Therefore, the RIS height and tilt angle must be optimized jointly.
Luckily, $\hRIS$ and $\theta_0 \in (0, \frac{\pi}{2} - \arccos{\frac{\yRIS}{d_{1}}})$ are both bounded.
Thus, given a grid of users, the optimal pair $(\hRIS^*, \theta_0^*)$ can be found by searching with $\xRIS^* = 0$ under different system layout parameters.

\addtolength{\topmargin}{-0.32in}

\section{Simulation results}
\label{sec:sim}

We will first illustrate snapshots of the system under several blockers showing the serving status of users and the rate heat maps.
Then we will verify the optimal RIS location by comparing both coverage ratio and average rate.
Next, we will obtain the optimal RIS deployment that maximizes the average rate, providing numerical results under different system layouts.
Finally, we will analyze the user rate performance and will compare it with the no RIS case.

\begin{figure}[t]
  \centering
  \subfigure[Serving status.]{
  \label{fig:servingstatus}
  \includegraphics[width= 3.3 in]{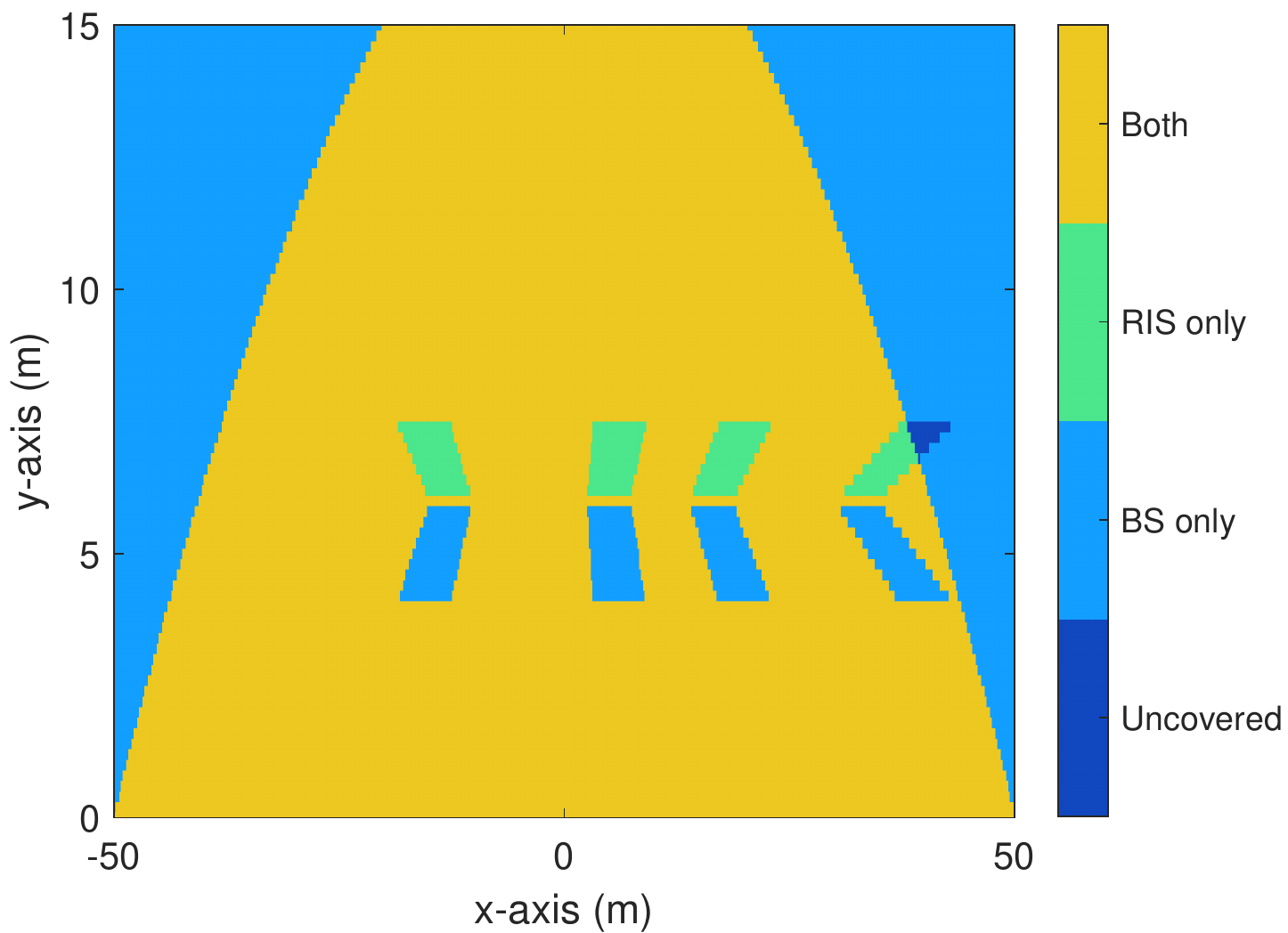}}
  \hspace{0em}
  \subfigure[Rate heat map.]{
  \label{fig:rateheatmap}
  \includegraphics[width= 3.2 in]{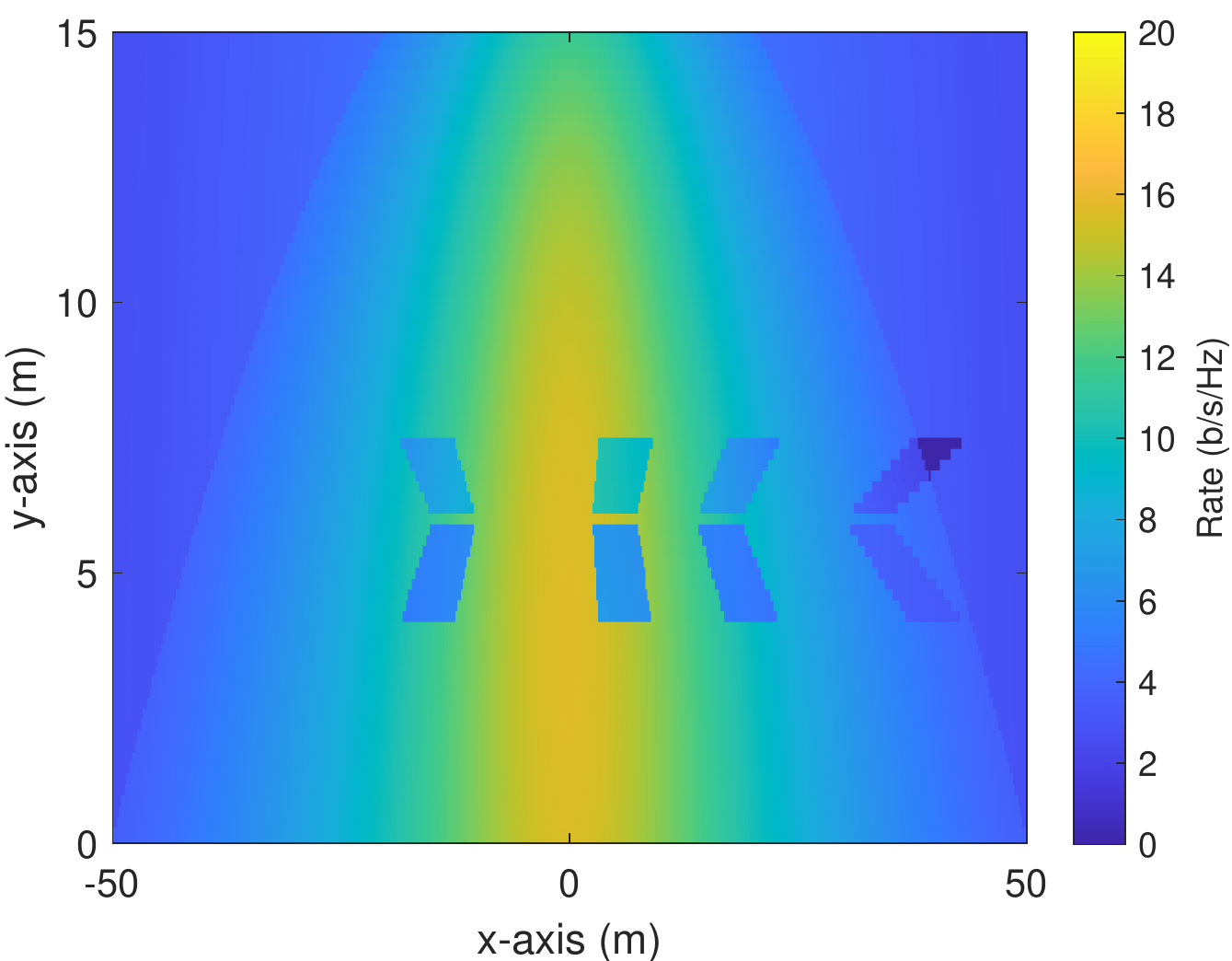}}
  \vspace{-0.0cm}
  \caption{Snapshot of one realization of the blockers, with $\yB = 6$ m, $\hBS = 10$ m, $\yRIS = 14$ m. A RIS of side length $0.5$ m is placed at $\hRIS=10$ m and tilted $30^o$ downwards. (a) Serving status. (b) Rate heat map.}
  \label{fig:Snapshots}
  \vspace{-0.4cm}
\end{figure}

The simulations are performed at a center frequency of 60 GHz. The antenna gains satisfy $\Gt\Gr = 100$.
To obtain the gain for one RIS's element we follow the general expression in \cite{TangEtAlWirelessCommunicationsReconfigurableIntelligent2021}. Thus, we define the normalized power radiation pattern of the $n$-th element of the RIS as $F(\theta_n, \phi_n) = \cos^3{\theta_n}$, $\theta_n \in [0, \frac{\pi}{2}]$, with elevation angle $\theta_n$ and azimuth angle $\phi_n$.
Under this definition, the gain of the $n$-th RIS element is $G=8$ \cite{TangEtAlWirelessCommunicationsReconfigurableIntelligent2021}.
In each iteration, the locations of the blockers $\xB$ are generated independently, and the number of blockers $M$ is generated by a Poisson distribution as in (\ref{eq:Poisson}), with $\lambdaB = 1$.
The resolution of the users' locations is set to $0.5$ m.
The threshold of path loss to support communication is set to $\gammath = 2.5\times10^{8}$, to make sure the BS covers the whole cell.
The height of BS and blockers is set to $\hBS = 10$ m and $\hB = 2$ m, and all the blockers are located on one lane with $\yB=6$ if not specified.
The widths of the road are $\yRIS = 14$ m, $22$ m and $30$ m, which are the approximate results based on $3.7$ m width per lane.
The BS coverage range in the x-axis is $R = 50$ m.
The length of the blocker is $L=4.8$ m.
The length and width of a RIS element is $a=b=\lambda/2$.
The number of elements is set to make sure the size of the RIS is $0.5$ m $\times 0.5$ m, i.e., $N=200 \times 200$.
Under this RIS parameter setting, the Fraunhofer distance is $400$ m, which makes the transmissions considered in our system be in the near-field.
In addition, the choice of this number of elements will make the BS-RIS-user link have an area-averaged rate comparable to that of the BS-user link, namely $4.96$ bps/Hz and $4.55$ bps/Hz, under the same setting as in Figs. \ref{fig:Snapshots}.
A higher number of RIS elements will result in a higher rate of the BS-RIS-user link, since the path loss is based on the summation of each element's contribution.
The RIS tilt angle $\theta_0$ is defined on a grid with a resolution of $1^o$.
The SNR is set as $90$ dB to compensate for the large-scale fading.

First, we show two snapshots of the serving status under blockages and the corresponding user rate heat maps.
Figs. \ref{fig:Snapshots} are generated with four blockers located at $\yB=6$ m and $\xB = -15$ m, $2.5$ m, $13.8$ m and $30$ m.
In Fig. \ref{fig:servingstatus}, different colors indicate different serving states.
Particularly, the yellow color indicates the region where users are served by both links, which is bounded by the RIS coverage rate with $\PLRIS \leq \gammath, \forall (x,y) \in B_0$ and (\ref{eq:PL_RIS}).
The green color indicates the region where BS-user link is blocked, and the user can only be served by the BS-RIS-user link.
Noticeably, there is a deep blue colored region where users are out of RIS coverage, and thus have no service.
The light blue region is where the users can be covered only by the BS-user link, due to either being out of RIS range or blockages.
With the rate for each user calculated by (\ref{eq:rateRISm}) and (\ref{eq:rateboth}) based on their serving status, the rate heat map for the considered region is shown in Fig. \ref{fig:rateheatmap}.
The center region has higher rates than the edge region because of the influence of elevation angles and link distances.
With RIS placed facing the BS and tilted optimally, most of the users are well-covered and have satisfactory rates.

\begin{figure}[t]
  \centering
  \subfigure[Coverage ratio.]{
  \label{fig:coverageratio}
  \includegraphics[width= 3.1 in]{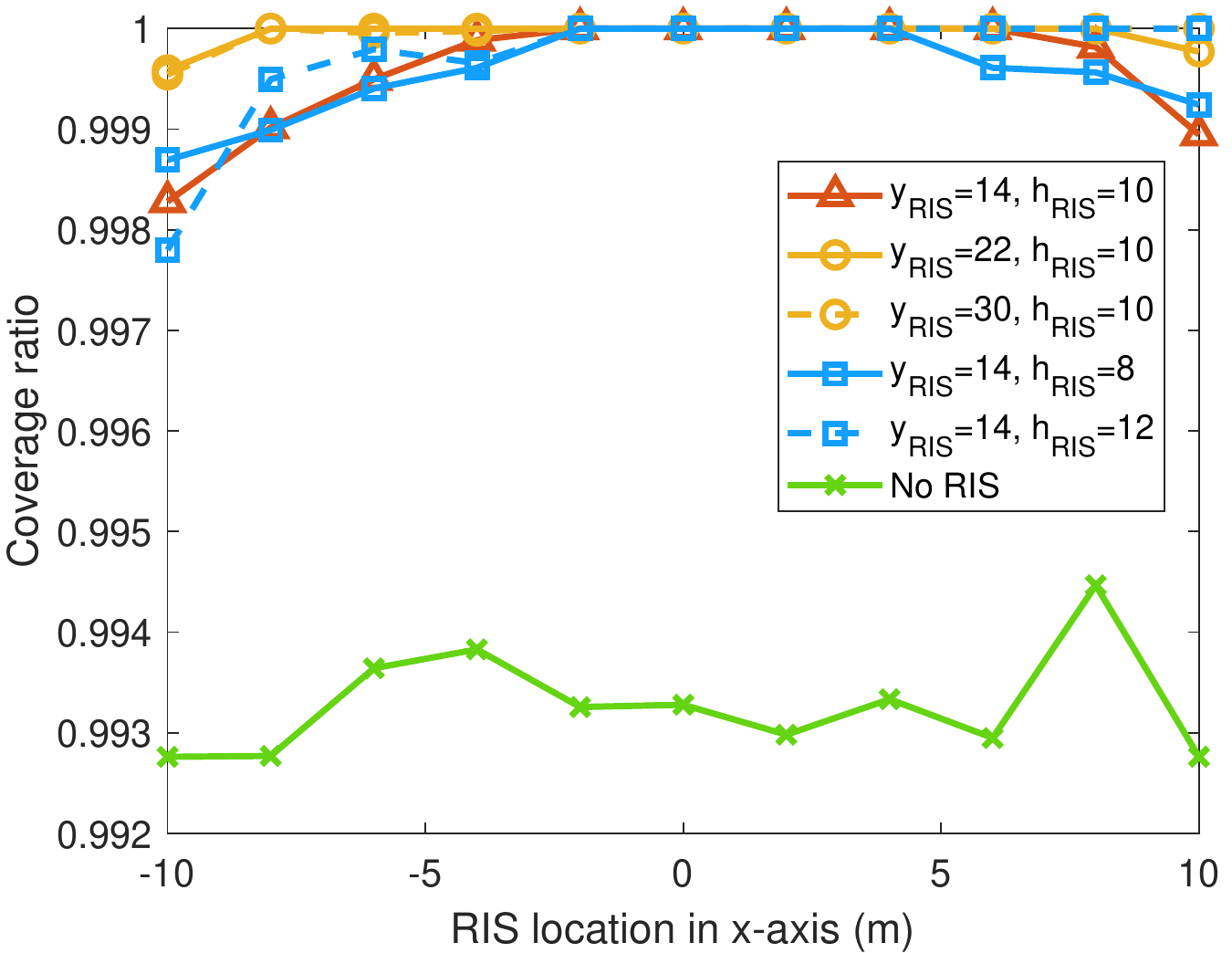}}
  \hspace{0em}
  \subfigure[Area-averaged rate]{
  \label{fig:areaaveragedrate}
  \includegraphics[width= 3.1 in]{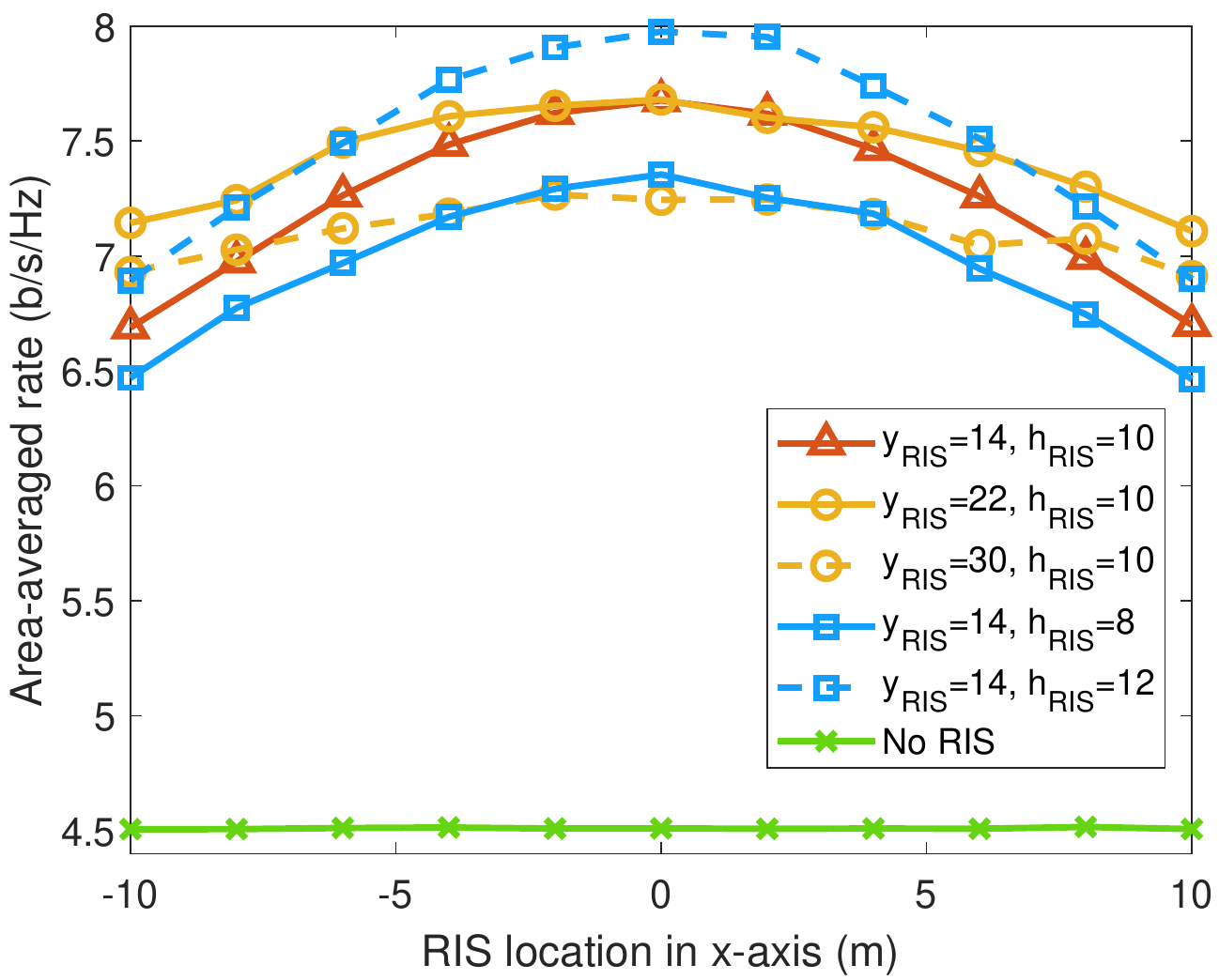}}
  \vspace{-0.0cm}
  \caption{Coverage ratio and corresponding area-averaged rate with different RIS locations $\xRIS$. The RIS tilt angle is optimal for every parameter setting.}
  \label{fig:findxRIS}
  \vspace{-0.1cm}
\end{figure}

We show in  Fig.~\ref{fig:findxRIS} the coverage ratio as defined in (\ref{eq:rho_N}), and the area-averaged rate as in (\ref{eq:R_N}) with different RIS locations $\xRIS$.
For comparison purposes, the results with no RIS are also shown.
It can be seen from Fig. \ref{fig:coverageratio} that, under our deployment strategy, where a RIS is deployed on the opposite side of the BS, the coverage ratio goes from $99\%$ to $99.9\%$, which helps the system to meet ultra reliability targets. 
The conclusion that the optimal RIS location in the x-axis is $\xRIS = 0$ as discussed in Sec. \ref{sec:opt_x_RIS} can be verified in Fig. \ref{fig:areaaveragedrate}.
The rate improvement by waterfilling-based power allocation ratio $\beta$ as in (\ref{eq:rateboth}), together with the coverage improvement due to introducing a RIS, having it optimally deployed can improve at least by $2.5$ b/s/Hz the area-averaged rate, which is over a $50\%$ improvement.

\begin{figure}[t]
  \centering
  \subfigure[Optimal tilt angle.]{
  \label{fig:opttiltangle}
  \includegraphics[width= 3.0 in]{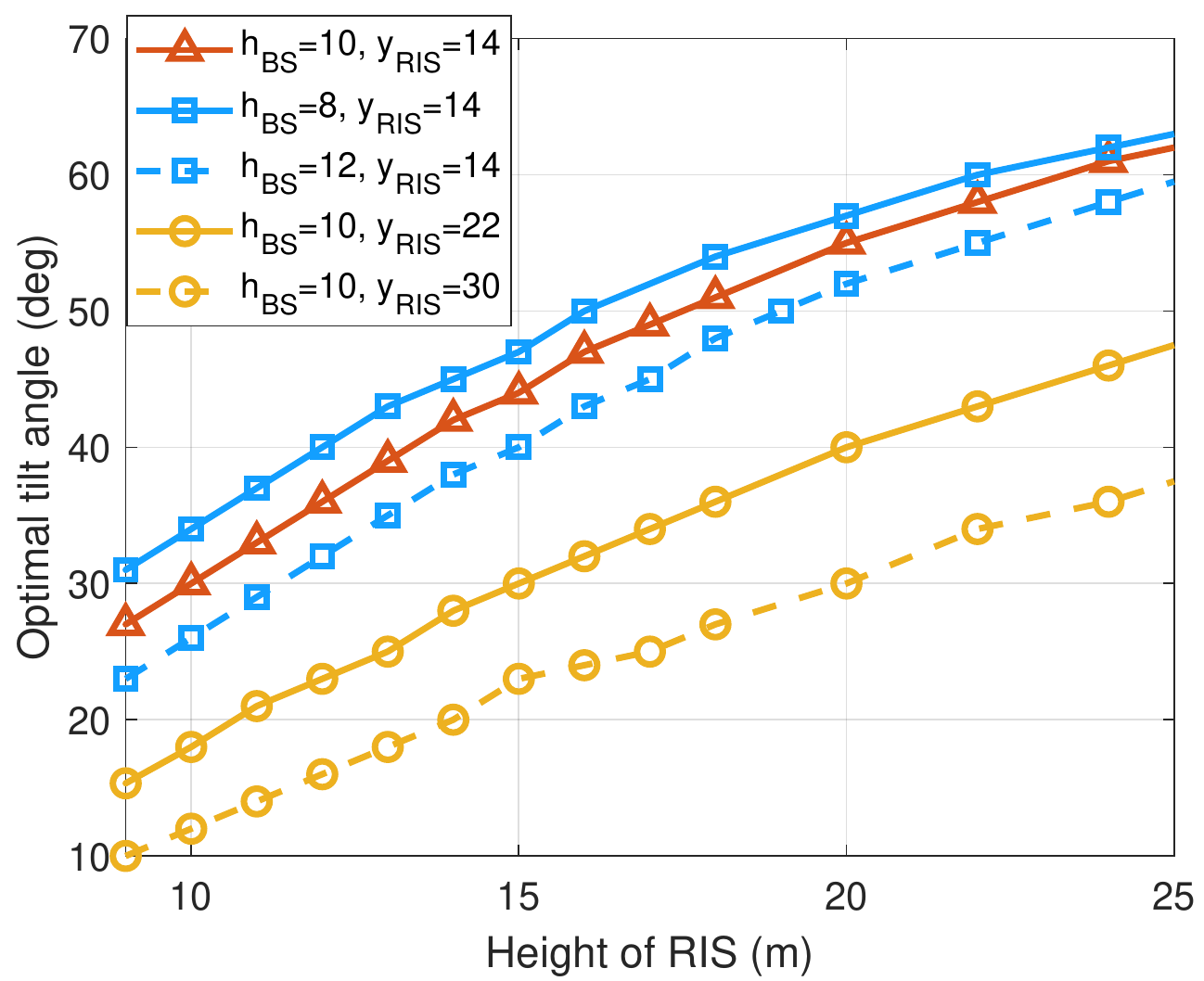}}
  \hspace{0em}
  \subfigure[Area-averaged rate under optimal tilt angle.]{
  \label{fig:optrate}
  \includegraphics[width= 3.0 in]{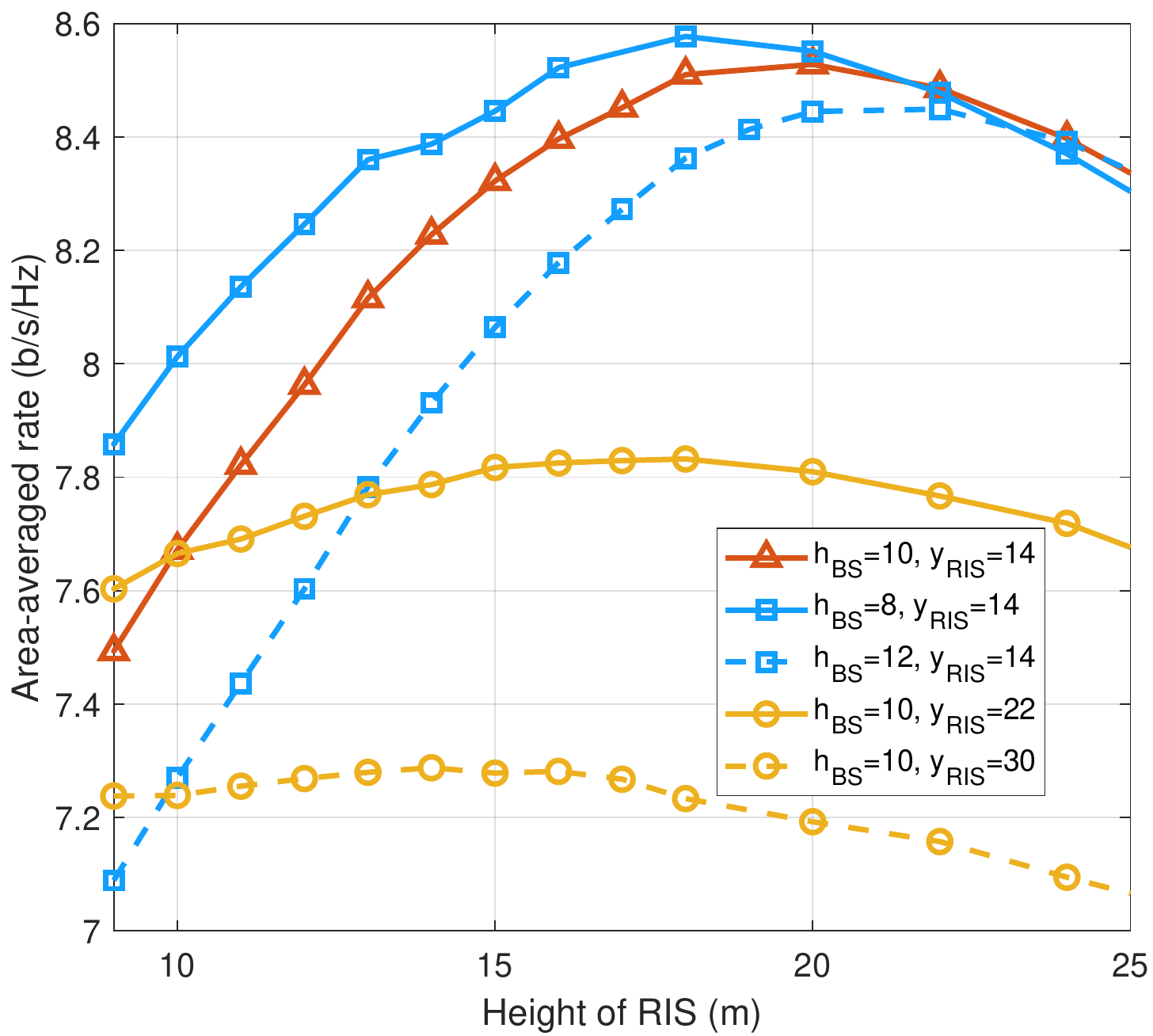}}
  \vspace{-0.0cm}
  \caption{Optimal tilt angle and the corresponding area-averaged rate versus height of RIS under optimal tilt angle.}
  \label{fig:optdeployment}
  \vspace{-0.5cm}
\end{figure}

Now we fix the RIS at the optimal location with $\xRIS^* = 0$, and try to find insights about the optimal deployment height $\hRIS$ and the downward tilt angle $\theta_0$.
As discussed in Sec. \ref{sec:opt_h_RIS_theta0}, given a RIS height $\hRIS$, the tilt angle is a trade-off between the BS-RIS elevation angle and the RIS-user elevation angle. The trend of the optimal RIS tilt angle and the corresponding area-averaged rate under optimal deployment are shown in Fig.~\ref{fig:optdeployment}.
Specifically, in Fig. \ref{fig:opttiltangle}, the optimal tilt angle increases as the height of RIS $\hRIS$ increases, which is for RIS to better ``see" the BS.
The effect of $\hBS$ on the tilt angle is that a higher BS requires a smaller tilt angle.
With a wider street (larger $\yRIS$), the required tilt angle is smaller.
The blocker's lane $\yB$ does not impact the optimal tilt angle, since it only affects the shaded area length, which can not be compensated by tilting the RIS.

The area-averaged rate under optimal RIS tilt angle as a function of $\hRIS$ is shown in Fig. \ref{fig:optrate}. A higher RIS can have a larger tilt angle to better serve the considered region, which explains the increase at first.
But limited by the large-scale attenuation, the RIS can not be unlimitedly high, which explains the decrease.
Other system layout parameters' effects can be found by comparing some of the lines in Fig. \ref{fig:optrate}.
The effect of $\hBS$ is reflected by comparing the red line with the blue lines and the conclusion is that a lower BS is beneficial when the height of BS and RIS are in the realistic range.
The effect of $\yRIS$ is reflected by comparing the red line with the black lines.
For a narrow street, a higher RIS will significantly improve the area-averaged rate while the specific deployment of the  RIS optimally has less impact on the rate performance when the street is wide.

\begin{figure}[t]
    \centering
    \includegraphics[width= 3.1 in]{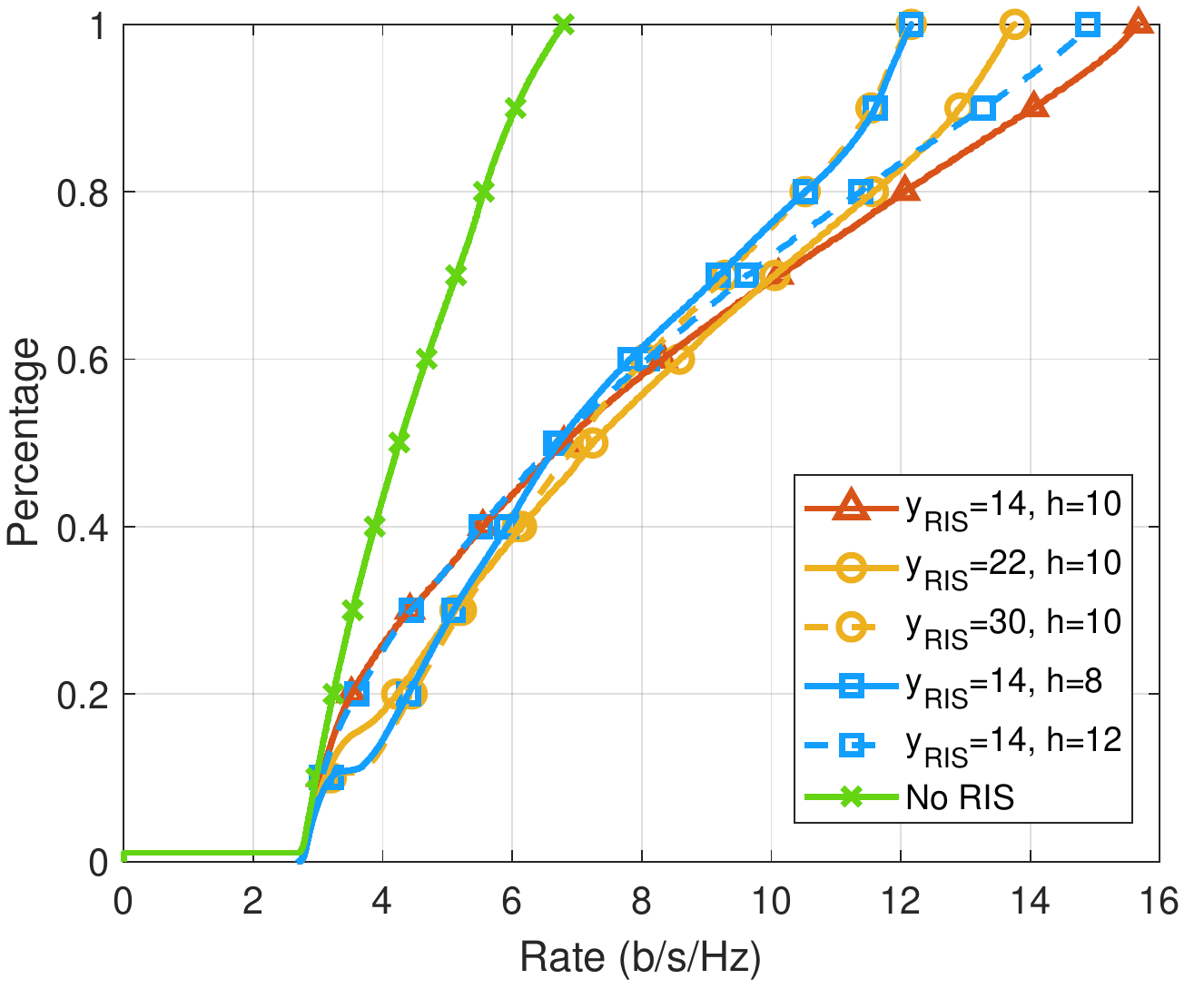}
    \vspace{-0.3cm}
    \caption{User rate distribution performance with optimal tilt angle.}
    \label{fig:cdf}
    \vspace{-0.2cm}
\end{figure}

Finally, in Fig. \ref{fig:cdf} we evaluate the user rate distribution performance and compare it with the same setting without RIS.
The height of BS and RIS are set to be equal for the ease of realistic deployment when they are mounted on the lamp pole, i.e., $\hBS=\hRIS=h$.
With the RIS optimally tilted, the improvement in user rate performance is significant.
Specifically, the percentage of users having no coverage is very close to $0$.
Furthermore, $50\%$ of the users can achieve a rate higher than approximately $7$ bps/Hz with an optimally deployed RIS, while that number is less than $4.5$ b/s/Hz without RIS. Another conclusion from this figure is that as long as the RIS is optimally deployed, the rate that $50\%$ users can get is almost the same, while the highest achievable rate is decided by the system parameters.

\section{Conclusions}

In this paper, we designed the optimal location, height and tilt angle of a RIS  in a 3D mmWave vehicular system under blockages.
The considered performance metrics are the coverage ratio and the area average rate, obtained with a near field beamforming model.
The resulting coverage ratio is close to $1$, and the area-average rate can be improved by at least $50\%$ under optimal RIS deployment.
The optimal RIS location in the x-axis is facing the BS, while the optimal tilt angle given a RIS height can be obtained through numerical search.

\bibliographystyle{IEEEtran}
\bibliography{xiaowen}

\end{document}